\documentclass[preprint,superscriptaddress,pra]{revtex4}
\def\<{\langle }
\def\>{\rangle }
\def\kk{\>\!\>}
\def\bb{\<\!\<}
\def\Tr{\hbox{Tr} }
\def\map#1{\mathscr{#1}}
\def\sH{\mathcal H}
\usepackage{amsmath,mathrsfs}
\usepackage{amssymb}
\usepackage{amsthm}
\usepackage{graphicx}
\usepackage{epsfig}
\usepackage{subfigure}
\begin{document}

\title{Superbroadcasting of continuous variables mixed states}

\author{Giacomo M. D'Ariano} 
\email{dariano@unipv.it}
\author{Paolo Perinotti}
\email{perinotti@fisicavolta.unipv.it}
\affiliation{Dipartimento di Fisica ``A. Volta'' and CNISM, via Bassi
  6, I-27100 Pavia, Italy.}  
\author{Massimiliano F. Sacchi}
\email{msacchi@unipv.it}
\affiliation{Dipartimento di Fisica ``A. Volta'' and CNISM, via Bassi
  6, I-27100 Pavia, Italy.}  \affiliation{CNR - Istituto Nazionale per
  la Fisica della Materia, Unit\`a di Pavia, Italy.}

\date{\today}

\begin{abstract}
  We consider the problem of broadcasting quantum information encoded
  in the average value of the field from $N$ to $M>N$ copies of mixed
  states of radiation modes. We derive the broadcasting map that
  preserves the complex amplitude, while optimally reducing the noise
  in conjugate quadratures. We find that from two input copies
  broadcasting is feasible, with the possibility of simultaneous
  purification ({\em superbroadcasting}). We prove similar results for
  purification ($M\leq N$) and for phase-conjugate broadcasting.
\end{abstract}

\maketitle

\section{Introduction}

Quantum cloning is impossible \cite{noclon}. This means that one
cannot produce a number of independent physical systems prepared in
identical states out of a smaller amount of systems prepared in the
same state. Since the formulation of the no-cloning theorem the search
for quantum devices that can perform cloning with the highest possible
fidelity gave rise to a whole branch in the literature. Optimal
cloners have been found, for qubits \cite{buzhill,gismass,bruss}, for
general finite-dimensional systems \cite{werner}, for restricted sets
of input states \cite{darlop,darmacc}, and for infinite-dimensional
systems such as harmonic oscillators---the so called continuous
variables cloners \cite{cerf}. However, for the case of mixed states,
a different type of cloning transformation can be considered---the
so-called {\em broadcasting}---in which the output copies are in a
globally correlated state whose local ``reduced'' states are identical
to the input states. This possibility has been considered in Ref.
\cite{nobro}, where it has been shown that broadcasting a single copy
from a noncommuting set of density matrices is always impossible.
Later, such a result has been considered in the literature as the
generalization of the no-cloning theorem to mixed states.  However,
more recently, for qubits an effect called {\em superbroadcasting}
\cite{prl} has been discovered, which consists in the possibility of
broadcasting the state while even increasing the purity of the local
state, for at least $N\ge 4$ input copies, and for sufficiently short
input Bloch vector (and even for $N=3$ input copies for
phase-covariant broadcasting instead of universal covariance
\cite{pra}).

In the present paper, we analyze the broadcasting of continuous
variable mixed states by a signal-preserving map. More precisely, this
means that we consider a set of states obtained by displacing a fixed
mixed state by a complex amplitude in the harmonic oscillator phase
space, while the broadcasting map is covariant with respect to the
(Weyl-Heisenberg) group of complex displacements. We will focus mainly
on displaced thermal states (which are equivalent to coherent states
that have suffered Gaussian noise), however, all results of the
present paper hold in terms of noise of conjugated quadratures for the
set of states obtained by displacing any fixed state.

As we will see, superbroadcasting is possible for continuous variable
mixed states, namely one can produce a larger number of copies, which
are purified locally on each use, and with the same signal of the
input.  For displaced thermal states, for example, superbroadcasting
can be achieved for at least $N=2$ input copies, with thermal photon
number $\overline{n}_{in}\ge\frac13$, whereas, for sufficiently large
$\overline{n}_{in}$ at the input, one can broadcast to an unbounded
number $M$ of output copies. For purification (i.e.  $M\le N$), quite
surprisingly the purification rate is
$\overline{n}_{out}/\overline{n}_{in}=N^{-1}$, independently on $M$.
The particular case of 2 to 1 for noisy coherent states has been
reported in Ref. \cite{ula}. We will prove also similar results for
broadcasting of phase-conjugated copies of the input.

The paper is organized as follows. In Section \ref{covar} we introduce
the problem of covariant broadcasting, deriving the general form of a
covariant channel (trace-preserving CP map), and introduce a special
channel that broadcast from $N$ to $M>N$ copies. In Section
\ref{optim} we prove that such a channel is optimal for broadcasting
any noisy displaced state. In Section \ref{puri} we consider the same
problem for purification (i.e. $M<N$). In Section \ref{phaconj} we
derive superbroadcasting for the output copies with a conjugate phase
with respect to the originals. In Section VI we show the optimality by
a simpler derivation, namely by exploiting the bounds from the theory
of linear amplification (which is then based on supplementary
assumptions). In Sec. VII we show a simple experimental scheme to
achieve optimal broadcasting/purification. Section \ref{concl} closes
the paper with a summary of results and some concluding remarks.

\section{Covariant broadcasting for the Weyl-Heisenberg group}
\label{covar}
We consider the problem of broadcasting $N$ input copies of displaced
(generally) mixed states of harmonic oscillators (with boson
annihilation operators denoted by $a_0,\ a_1,...,a_{N-1}$) to $M$
output copies (with boson annihilation operators $b_0,\
b_1,...,b_{M-1}$). In order to preserve the signal, the broadcasting
map $\map B$ must be covariant, i.~e. in formula
\begin{equation}
  \map B(D(\alpha)^{\otimes N}\Xi D(\alpha)^{\dag\otimes N})=D(\alpha)^{\otimes M}\map B(\Xi)D(\alpha)^{\dag\otimes M},
\label{cov}
\end{equation}
where $D_c(\alpha)=\exp(\alpha c^\dag-\alpha^* c)$ denotes the
displacement operator, and $\Xi$ represents an arbitrary $N$-partite
state. It is useful to consider the Choi-Jamio\l kowski bijective
correspondence of completely positive (CP) maps $\map B$ from
$\sH_\mathrm{in}$ to $\sH_\mathrm{out}$ and positive operators $
R_{\map B}$ acting on $\sH_\mathrm{out}\otimes\sH_\mathrm{in}$, which
is given by the following expressions
\begin{equation}
\begin{split}
  &R_{\map B}=\map B\otimes \map I(|\Omega\>\<\Omega|)\;,\\
  &\map B(\rho)=\Tr_\mathrm{in}[(I_\mathrm{out}\otimes\rho^\tau )R_{\map
    B}]\;,
\end{split}
\end{equation}
where $|\Omega\>=\sum_{n=0}^{\infty}|\psi_n\>|\psi_n\>$ is a maximally
entangled vector of $\sH_\mathrm{in} ^{\otimes 2}$, and $X^\tau $ denotes
transposition of $X$ in the basis $|\psi_n\>$. In terms of the
operator $R_{\map B}$ the covariance property \eqref{cov} can be
written as
\begin{eqnarray} 
[R_{\map B},D(\alpha )^{\otimes M}\otimes D(\alpha
  ^*)^{\otimes N}]=0\;,\qquad \forall \alpha \in {\mathbb
    C}\;.\label{com}
\end{eqnarray}
In order to deal with this constraint we introduce the multisplitter
operators $U_a$ and $U_b$, that perform the unitary transformations
\begin{eqnarray}
  && U_a a_k U^\dag _a =\frac 1{\sqrt N}\sum_{l=0}^{N-1}e^{\frac{2\pi i kl}N} a_{l} \;,
  \nonumber \\& & 
  U_b b_k U^\dag _b =\frac 1{\sqrt M} \sum_{l=0}^{M-1}e^{\frac{2\pi i
      kl}M}b_l
  \;.\label{multisplit}
\end{eqnarray}
Notice that such transformations perform a Fourier transform over all
input and output modes.   
Moreover, we will make use of the squeezing transformation
$S_{a_0b_0}$ defined as follows
\begin{equation}
\begin{split}
  & [S_{a_0 b_0},a_n]=[S_{a_0 b_0},b_n]=0,\; n>0\\
  & S_{a_0 b_0}a_0^\dag S_{a_0b_0}^\dag =\mu a_0^\dag -\nu b_0 \;,\\&
  S_{a_0b_0}b_0S_{a_0b_0}^\dag =\mu b_0 -\nu a_0^\dag\;,
\end{split}
\label{ampli}
\end{equation}
with $\mu = \sqrt{M/(M-N)}$ and $\nu =\sqrt{N/(M-N)}$. The squeezing
transformation here acts just as an hyperbolic transformation for just
modes $a_0$ and $b_0$, by leaving all other modes unaffected. In terms of
such operators, condition (\ref{com}) becomes
\begin{eqnarray}
[S^\dag _{a_0b_0}(U^\dag_b\otimes U^\dag _a) R_{\map B} (U_b\otimes U_a) S_{a_0b_0},
  D_{b_0}(\sqrt{M-N} \alpha )]=0\;.\label{commrel}
\end{eqnarray}
Hence, upon introducing an operator $B$ on modes
$b_1,...,b_{M-1},a_0,...,a_{N-1}$, the operator $R_{\map B}$ can be
written in the form
\begin{eqnarray}
R_{\map B}= (U_b\otimes U_a) S_{a_0b_0} (I_{b_0}\otimes B)
S^\dag _{a_0b_0}(U^\dag_b\otimes U^\dag _a).
\end{eqnarray}
Notice that $R_{\map B}\geq 0$ is equivalent to $B\geq 0$. The further
condition that $\map B$ is trace-preserving in terms of $R_{\map B}$
becomes $\Tr_{b}[R_{\map B}]=I_{a}$, $b$ and $a$ collectively denoting
all output and input modes, respectively. From the trace and
completeness relations for the set of displacement operators,
namely $\int d^2 \alpha \,D(\alpha )A D^\dag (\alpha )=\hbox{Tr}[A]I$,
and $A=\int d^2 \alpha \,\Tr[D^\dag (\alpha )A]D(\alpha )$, (see,
e.g., Ref. \cite{bm}), the condition $\Tr_{b}[R_{\map B}]=I_{a}$ is verified
iff
\begin{eqnarray}
\left(\prod _{i=0}^{M-1} \int d^2 \beta _i \right ) 
\left(\bigotimes _{i=0}^{M-1} D_{b_i}(\beta _i ) \right)S_{a_0b_0}(I_{b_0} \otimes B)S^\dag _{a_0b_0}
\left(\bigotimes _{i=0}^{M-1} D^\dag _{b_i}(\beta _i)\right)= I
\;.\label{tracep}
\end{eqnarray}
From the relation $D_{b_0}(\beta _0)S_{a_0b_0}=S_{a_0b_0}D_{b_0}(\mu
\beta _0)
D_{a_0}^\dag (\nu \beta _0)$, one obtains the condition
\begin{eqnarray}
  \hbox{Tr}_{b/b_0,a_0}[B]=\nu ^2
  I_{a/a_0}\;,
\label{trb}
\end{eqnarray}
where $a/a_i$ denote all the input modes apart from $a_i$, and
similarly for $b/b_i$.

We will now consider the map corresponding to
\begin{equation}
  B=\nu ^2 \,
  |0\>\<0|_{b/b_0}\otimes|0\>\<0|_{a_0}\otimes
  I_{a/a_0}\;.
\label{bcho}
\end{equation}
Applying the corresponding map $\map B$ to a generic $N$-partite state
$\Xi$ we get
\begin{equation}\label{optmap}
  \map B(\Xi)=\Tr_{a}[(I_b\otimes\Xi^\tau )(U_b\otimes U_a) S_{a_0b_0} (I_{b_0}\otimes B)
  S^\dag _{a_0b_0}(U^\dag _b\otimes U^\dag_a)]\,,
\end{equation}
which is equivalent to
\begin{equation}
  \map B(\Xi)=\Tr_{a}[(I_b\otimes U^\dag _a\Xi^\tau U_a) (U_b\otimes I_a) S_{a_0b_0} (I_{b_0}\otimes B)
  S^\dag _{a_0b_0}(U^\dag_b\otimes I_a)]\,.
\end{equation}
Using the expression in Eq.~\eqref{bcho} we obtain
\begin{equation}
  \map B(\Xi)=U_b\left\{\Tr_{a_0}[(I_{b_0}\otimes\xi^\tau _{a_0}) S_{a_0b_0} (I_{b_0}\otimes |0\>\<0|_{a_0})
    S^\dag _{a_0b_0}]\otimes|0\>\<0|_{b/b_0}\right\}U^\dag_b\,,\label{13}
\end{equation}
where $\xi^\tau =\Tr_{a/a_0}[U^\dag _a\Xi^\tau U_a]$. Notice that
\begin{equation}
\begin{split}
  \xi=&\int\frac{d^2\gamma}\pi
  D(\gamma)^\tau \Tr[(D_{a_0}(\gamma)^\dag\otimes I_{a/a_0})U_a^\dag\Xi^\tau 
  U_a]\\=&\int\frac{d^2\gamma}\pi
  D(\gamma)^\tau \Tr[U^*_a(D_{a_0}(\gamma)^*\otimes I_{a/a_0})U_a^\tau \Xi],
\end{split}
\end{equation}
and taking the complex conjugate of Eq.~\eqref{multisplit} we have
\begin{equation}
\begin{split}
  \xi=&\int\frac{d^2\gamma}\pi D(\gamma)^\tau \Tr[D(\gamma^*/\sqrt{N})^{\otimes N} \Xi ]\\
  =&\int\frac{d^2\gamma}\pi D(\gamma)^\tau \Tr[(D_{a_0}(\gamma)^*\otimes
  I_{a/a_0}) U_a^\dag\Xi U_a]=\Tr_{a/a_0}[U_a^\dag\Xi U_a]\,
\end{split}
\end{equation}
Now, we can easily evaluate $S_{a_0b_0} (I_{b_0}\otimes
|0\>\<0|_{a_0}) S^\dag _{a_0b_0}$, by expanding the vacuum state as
\begin{equation}
  |0\>\<0|_{a_0}=\int\frac{d^2\gamma}\pi 
  e^{-\frac{|\gamma|^2}2}D_{a_0}(\gamma)\,,
\end{equation}
obtaining
\begin{equation}
  S_{a_0b_0} (I_{b_0}\otimes |0\>\<0|_{a_0}) S^\dag _{a_0b_0}=\int\frac{\nu^2d^2\gamma}{\pi} e^{-\frac{|\gamma|^2}2}D_{b_0}(\nu\gamma^*)\otimes D_{a_0}(\mu\gamma)\,.
\end{equation}
Hence, Eq. (\ref{13}) can be rewritten as
\begin{equation}
  \map B(\Xi)=\int\frac{d^2\gamma}\pi U_b (D_{b_0}
  (\gamma^*)\otimes|0\>\<0|_{b/b_0})U_b^\dag e^{-\frac{|\gamma|^2}{2\nu^2}}\Tr[D_{a_0}(\mu\gamma/\nu)\xi^\tau ]\,.
\end{equation}
As an example, we will now consider $N$ displaced thermal states
\begin{equation}
  \rho_\alpha\doteq\frac1{\bar n+1}D(\alpha)\left(\frac{\bar n}{\bar n+1}\right)^{a^\dag a}D(\alpha)^\dag\,,
\label{therm}
\end{equation}
from which we want to obtain $M$ states, the purest as possible.
Thanks to the covariance property, it is sufficient to focus attention
on the output of $\rho_0^{\otimes N}$. For a tensor product of thermal
input states $\Xi=\rho_0^{\otimes N}$, exploiting the fact that
$U_a^\dag (\sum_{j=0}^{N-1}a^\dag_j a_j) U_a=\sum_{j=0}^{N-1}a^\dag_j
a_j$, we have
\begin{equation}
  \xi=\xi^\tau =\rho_0\,,
\end{equation}
and recalling the following expression for the thermal states
\begin{equation}
  \frac1{\bar n+1}\left(\frac{\bar n}{\bar n+1}\right)^{a^\dag a}=\int\frac{d^2\beta}\pi e^{-\frac{|\beta|^2}2(2\bar n+1)}D(\beta)\,,
\end{equation}
we obtain 
\begin{equation}
\begin{split}
  \map B\left(\rho_0^{\otimes N}\right)=&\int\frac{d^2\gamma}\pi U_b
  (D_{b_0} (-\gamma^*)\otimes|0\>\<0|_{b/b_0})U_b^\dag
  e^{-\frac{|\gamma|^2}{2\nu^2}[\mu^2(2\bar n+1)+1]}\\
  &=\int\frac{d^2\gamma}{\pi\bar n'} U_b
  (|\gamma\>\<\gamma|_{b_0}\otimes|0\>\<0|_{b/b_0})U_b^\dag
  e^{-\frac{|\gamma|^2}{\bar n'}} \\
  &=\int\frac{d^2\gamma}{\pi\bar n'}
  |\gamma/\sqrt{M}\>\<\gamma/\sqrt{M}|^{\otimes M}
  e^{-\frac{|\gamma|^2}{2\bar n'}} = \int\frac{M d^2\gamma}{\pi\bar
    n'} |\gamma\>\<\gamma|^{\otimes M} e^{-\frac{M|\gamma|^2}{\bar
      n'}}
\label{finalnumb}
,
\end{split}
\end{equation}
where
\begin{equation}
  2\bar n'+1=\frac1{\nu^2}\left[\mu^2(2\bar n+1)+1\right]=\frac{2M\bar n+2M-N}{N}.
\label{finalnumb2}
\end{equation}
The above state is permutation-invariant and separable, with thermal
local state at each mode with average thermal photon
\begin{equation}
  \bar n''=\frac{\bar n'}{M}=\frac{M\bar n+M-N}{MN}\,.
\label{super}
\end{equation}
More generally, for any state $\Xi$, the choice (\ref{bcho}) gives $M$
identical clones whose state can be written as
\begin{eqnarray}
  \rho ' = \int \frac{d^2 \alpha }{\pi}\,e ^{-\frac{|\alpha |^2}{2}(\frac
    1N -\frac 2M +1)}\, 
  \{\Tr [\Xi D^\dag (\alpha  /N)^{\otimes N}] \} \, D(\alpha )
  \;.\label{rhogen}
\end{eqnarray}
Since for any mode $c$ one has
\begin{eqnarray}
  \Delta x_c ^2 +\Delta y_c ^2  = \frac 12 + \< c^\dag c \> -|\<c\>|^2,
\label{anym}
\end{eqnarray}
it is easy to verify that the superbroadcasting condition (output
total noise in conjugate quadratures smaller than the input one), is
equivalent to require smaller photon number at the output than at the
input, namely
\begin{equation}
  \bar n\geq\frac{M\bar n+M-N}{MN}\,
  \quad\Leftrightarrow \quad \bar n\geq\frac{M-N}{M(N-1)}\,.
\end{equation}
This can be true for any $N>1$, and to any $M\le\infty$, since
\begin{equation}
  \lim_{M\to\infty}\frac{M-N}{M(N-1)}=\frac1{N-1}>0\,.
\end{equation}

\section{Proof of optimality for the channel in Eq. (\ref{optmap})}
\label{optim}
Actually, the solution given in Eq. (\ref{super}) is optimal. To prove
this, in the following we will show that the expectation of the total
number of photons $\Tr[\sum _{l=0}^{M-1} b_l^\dag b_l \map
B(\rho_0^{\otimes N})]$ of the $M$ clones of $\rho$ cannot be smaller
than $M \bar n''$. Since the multisplitter preserves the total number
of photons we have to consider the trace
\begin{eqnarray}
W \doteq \Tr\left [\left(\sum _{l=0}^{M-1} b_l^\dag b_l
\otimes  (U^\dag_a \rho_0^{\tau\otimes N}U_a) \right) S_{a_0b_0}(I_{b_0}\otimes B) S^\dag _{a_0 b_0}\right].\label{vutot}
\end{eqnarray}
We can write $W =W_0 + \sum _{l=1}^{M-1} W_l$, with
\begin{eqnarray}
  &&W_0 \doteq \Tr\left [S^\dag _{a_0 b_0} \left((b_0^\dag b_0\otimes I_{b/b_0})
      \otimes (U_a^\dag\rho_0^{\tau\otimes N}U_a)\right) S_{a_0b_0}
    (I_{b_0}\otimes B) \right],\nonumber\\
  &&W_l \doteq 
  \Tr\left [S^\dag _{a_0 b_0}\left((I_{b/b_l}\otimes b^\dag _l b_l )\otimes 
      (U_a^\dag\rho_0^{\tau\otimes N}U_a)\right) S_{a_0b_0}
     (I_{b_0}\otimes B)
  \right],\label{vuzel}
\end{eqnarray}
for $\ 1\leq l\leq M-1$. Now, since $W_l\geq0$, $W\geq W_0$.
Moreover, using the identity $c^\dag c = -
\partial _{\alpha \alpha ^*}e^{\frac{|\alpha |^2}{2}}D_c( \alpha ) |
_{\alpha =\alpha ^*=0}$, one obtains
\begin{eqnarray}
  &&\Tr_{b_0} \left [ S^\dag _{a_0 b_0} \left(b^\dag _0 b_0
      \otimes \sigma\right) S_{a_0b_0} \right] \nonumber \\& & 
  =- \partial _{\alpha \alpha^* }\int \frac {d^2
    \gamma }{\pi } \Tr _{b_0} [D_{b_0}(\mu \alpha - \nu \gamma ^*)\otimes
  D_{a_0}(\mu \gamma -\nu \alpha ^*)]\, \Tr[D(\gamma)^\dag\sigma]e^{\frac {|\alpha |^2}{2}} |_{\alpha =
    \alpha ^*=0}
  \nonumber \\& & =\left.
    - \frac {1}{\nu ^2}
    \partial _{\alpha \alpha ^*} e^{-\frac {|\alpha |^2}{\nu ^2}}\,e^{\frac {\alpha ^*}{\nu }a_0 ^\dag }\,
    e^{-\frac {\alpha }{\nu }a_0 } \Tr\left[e^{\frac {\mu\alpha ^*}{\nu }a_0 ^\dag }\,
      e^{-\frac {\mu\alpha }{\nu }a_0 }\sigma\right] \right|_{\alpha =\alpha ^*=0} 
  =\frac {a_0^\dag a_0 +\mu ^2\Tr[a^\dag_0a_0\sigma] +1}{\nu ^4},\label{squiquiz}
\end{eqnarray}
then, from Eq.~\eqref{trb} and positivity of $B$, one has
\begin{equation}
\begin{split}
  W_0=&\frac{\Tr [(I_{b/b_0} \otimes a^\dag_0a_0\otimes I_{a/a_0})
    \{I_{b/b_0}\otimes(U_a\rho _0 ^{\otimes N}U_a^\dag)^\tau\}
    B]}{\nu^4}\\
  &\frac{\mu^2\Tr[(I_{b/b_0}\otimes I_{a_0}\otimes \Tr_{a_0}[a^\dag_0
    a_0 (U_a\rho_0^{\otimes N}U_a^\dag)^\tau])B] +\nu^2}{\nu ^4}\\
  &\geq \frac{\mu ^2\bar n +1}{\nu ^2} = \frac NM \bar n +\frac
  {M-N}{N}=M \bar n''.\label{bnd}
\end{split}
\end{equation}
In fact, one can easily check that the choice of $B$ in Eq.
(\ref{bcho}) saturates the bound (\ref{bnd}).
\par Also the more general solution given in Eq. (\ref{rhogen}) is
optimal, in the sense that it represents the state of $M$ identical
clones with minimal photon number, which is given by
\begin{eqnarray}
  \Tr [b^\dag b \rho ' ]= \frac {\Tr [
    a^\dag a \rho _0 ]}{N}+\frac 1N -\frac 1M
  \;.
\end{eqnarray}
Notice that for $\bar n=0$ one has $N$ coherent states at the input,
and $\bar n''=\frac{M-N}{MN}$, namely one recovers the optimal cloning
for coherent states of Ref. \cite{cerfbrauns}.

From Eq. (\ref{anym}), one can see that our optimization maximally
reduces the total noise in conjugate quadratures. Alternatively, one
might minimize the output entropy, which would be informationally more
satisfactory. This case, however, turns out to be a non trivial task,
and is beyond the scope of this article.

\section{Purification}
\label{puri}
For $M<N$ one can look for the optimal ``purification'' map with $M$
output systems. The result can be obtained as in section \ref{covar},
provided that we replace the operator $S_{a_0b_0}$ in
Eq.~\eqref{ampli} with
\begin{equation}
\begin{split}
  & [T_{a_0 b_0},a_n]=[T_{a_0 b_0},b_n]=0,\; n>0\\
  & T_{a_0 b_0}a_0 T_{a_0b_0}^\dag =\mu a_0-\nu b_0^\dag \;, \\&
  T_{a_0b_0}b_0^\dag T_{a_0b_0}^\dag =\mu b_0^\dag -\nu
  a_0\;,\label{deamp}
\end{split}
\end{equation}
where now $\mu=\sqrt{\frac{N}{N-M}}$ and $\nu=\sqrt{\frac{M}{N-M}}$,
and the constraint in Eq.~\eqref{commrel} with
\begin{eqnarray} 
[T^\dag _{a_0b_0}(U^\dag_b\otimes U^\dag _a) R_{\map
    B} (U_b\otimes U_a) T_{a_0b_0}, D_{a_0}(\sqrt{N-M} \alpha
  )]=0\;,\label{commdea}
\end{eqnarray}
for all $\alpha $. Consequently, $R_{\map B}$ has the form
\begin{equation}
  R_{\map B}=(U_b\otimes U_a)T_{a_0b_0}(I_{a_0}\otimes B)T^\dag_{a_0b_0}(U_b^\dag\otimes U_a^\dag),
\end{equation}
and trace preservation is equivalent to
\begin{eqnarray}
  \Tr_{b}[B]=\mu ^2 I_{a/a_0}.
\label{trbde}
\end{eqnarray}
Now, we consider the map with
\begin{equation}
  B=\mu ^2 \,
  |0\>\<0|_{b}\otimes I_{a/a_0}.
\label{bchode}
\end{equation}
The corresponding output for given input state $\Xi$ is given by
\begin{equation}
  \map B(\Xi)=\int\frac{d^2\gamma}\pi U_b (D_{b_0}
  (\gamma)\otimes|0\>\<0|_{b/b_0})U_b^\dag e^{-\frac{|\gamma|^2}{2\mu^2}}\Tr[D_{a_0}(-\nu\gamma^*/\mu)\xi^\tau ],
\end{equation}
where $\xi=\Tr_{a/a_0}[U_a\Xi U_a^\dag]$. For $\Xi=\rho_0^{\otimes N}$
we have $\xi=\rho_0$, and
\begin{equation}
  \map B(\rho_0^{\otimes N})=\int\frac{d^2\gamma}{\pi}e^{-\frac{|\gamma|^2}{2\mu^2}[\nu^2(2\bar n+1)+1]}U_b(D_{b_0}(\gamma)\otimes|0\>\<0|_{b/b_0})U_b^\dag.
\end{equation}
The integral gives a thermal state for the mode $b_0$ with average
photon number $\bar n'$ such that $2\bar n'+1=\frac{\nu^2(2\bar
  n+1)+1}{\mu^2}=2\frac{M}{N}\bar n+1$, namely $\bar n'=\frac MN\bar
n$. Finally, one has
\begin{equation}
  \map B(\rho_0^{\otimes N})=\int\frac{Md^2\gamma}{\bar n'\pi}
  \,e^{-\frac{M|\gamma|^2}{\bar n'}}|\gamma\>\<\gamma|^{\otimes M}\;.
\end{equation}
Hence, the single-site reduced state is a thermal state with a
number of thermal photons 
\begin{equation}
\bar n''=\frac{\bar n}N\;,
\label{ennesec}
\end{equation}
which is rescaled with respect to the input by a factor $N$,
independently of the number of output copies. The same analysis as in
section \ref{optim} shows that this is the minimum output number
compatible with complete positivity of the map $\map B$, and then is
optimal.\par

For a generic input state $\Xi$ the local output state is given by
\begin{equation}
  \rho ' = \int \frac{d^2 \alpha }{\pi}\,e ^{-\frac{|\alpha |^2}{2}(1-\frac
    1N)}\, 
  \{\Tr [\Xi D^\dag (\alpha  /N)^{\otimes N}] \} \, D(\alpha ),
\end{equation}\par

Notice that both Eq.~\eqref{finalnumb2} and Eq.~\eqref{ennesec} give
$\bar n''=\frac{\bar n}N$ also for $M=N$, 
and this result can be proved as follows. The difference from the
previous proof resides in the fact that the squeezing operator
$S_{a_0b_0}$ is ill defined in this case. However, once we unitarily
transform $D(\alpha)^{\otimes N}\otimes D(\alpha)^{*\otimes N}$ to
$D_{b_0}(\sqrt N\alpha)\otimes I_{b/b_0}\otimes D(\sqrt
N\alpha)^*_{a_0}\otimes I_{a/a_0}$, the squeezing operator on modes
$a_0$ and $b_0$ is not needed, and it is sufficient to remark that the
representation $D_{b_0}(\sqrt N\alpha)\otimes D_{a_0}(\sqrt
N\alpha)^*$ is abelian, and its joint eigenvectors can be written as
\begin{equation}
|D(\beta)\kk\doteq\sum_{m,n=0}^{\infty}\<m|D(\beta)|n\>\ |m\>_{b_0}|n\>_{b_0}.
\end{equation}
Consequently, the covariance condition for the map $\map B$ is given
by
\begin{equation}
R_{\map B}=(U_b\otimes U_a)\int\frac{d^2\gamma}\pi|D(\gamma)\kk\bb D(\gamma)|_{a_0b_0}\otimes \Delta_{a/a_0,b/b_0}(\gamma) (U_b^\dag\otimes U_a^\dag),
\end{equation}
with the trace-preserving constraint expressed by
\begin{equation}
\int\frac{d^2\gamma}\pi \Tr_{b/b_0}[\Delta_{a/a_0,b/b_0}(\gamma)]=I_{a/a_0}.
\end{equation}
We consider the following form for $\Delta_{a/a_0,b/b_0}(\gamma)$
\begin{equation}
\Delta_{a/a_0,b/b_0}(\gamma)=\pi\delta^2(\gamma)I_{a/a_0}\otimes |0\>\<0|_{b/b_0},
\end{equation}
which gives
\begin{equation}
R_{\map B}=(U_b\otimes U_a)|I\kk\bb I|_{a_0b_0}\otimes |0\>\<0|_{b/b_0}\otimes I_{a/a_0} (U_b\otimes U_a)^\dag,
\end{equation}
and then we can prove optimality by the same technique used in the
other cases. The output of $\Xi$ is given by
\begin{equation}
\map B(\Xi)=U_b (\xi^\tau _{b_0}\otimes|0\>\<0|_{b/b_0})U_b^\dag,
\end{equation}
where $\xi=\Tr_{a/a_0}[U_a^\dag\Xi U_a]$, and for thermal states
$\Xi=\rho_0^{\otimes N}$ we have $\xi=\rho_0$ and
\begin{equation}
\map B(\rho_0^{\otimes N})=\int\frac{Nd^2\gamma}{\pi\bar n} e^{-\frac{N|\gamma|^2}{\bar n}}|\gamma\>\<\gamma|^{\otimes N},
\end{equation}
which is separable, and its local states are thermal states with
\begin{equation}
\bar n''=\frac {\bar n}N.
\end{equation}

\section{Phase-conjugating broadcasting}
\label{phaconj}
We now consider the problem of broadcasting with simultaneous
phase-conjugate output. This means that we look for the optimal
transformation where the average of the output field of each copy is
the complex conjugate with respect to the value of the input one. 
The
covariance property of such a map is the following
\begin{equation}
\map C(D(\alpha)^{\otimes N}\Xi D(\alpha)^{\dag\otimes N})=D(\alpha)^{*\otimes M}\map C(\Xi)D(\alpha)^{T\otimes M},
\end{equation}
for all $\alpha $, and in terms of $R_{\map C}$ this corresponds to
\begin{equation}
[D(\alpha)^{*\otimes(M+N)},R_{\map C}]=0.
\label{comconj}
\end{equation}
We will use the same multisplitters defined in Eq.~\eqref{multisplit},
and introduce the following beam-splitter
\begin{equation}
\begin{split}
& [U_{a_0 b_0},a_n]=[U_{a_0 b_0},b_n]=0,\; n>0\\
& U_{a_0b_0}b_0 U^\dag_{a_0b_0}=\eta b_0+\theta a_0\\
& U_{a_0b_0}a_0 U^\dag_{a_0b_0}=-\theta b_0+\eta a_0,
\end{split}
\end{equation}
with $\eta=\sqrt{\frac{M}{M+N}}$ and $\theta=\sqrt{\frac{N}{M+N}}$.
The covariance relation in Eq.~\eqref{comconj} can be written
\begin{equation} 
[U^\dag_{a_0b_0}(U^\dag_b\otimes U^\dag_a) R_{\map C}
(U_b\otimes U_a)U_{a_0b_0},D_{b_0}(\sqrt{M+N}\alpha)^*]=0.
\end{equation}
Analogously to the previous sections, the covariance condition translates
in the following form for $R_{\map C}$:
\begin{equation}
R_{\map C}= U_bU_a U_{a_0b_0} (I_{b_0}\otimes C)
U^\dag _{a_0b_0}U^\dag_bU^\dag _a,
\end{equation}
where $C$ is an operator on modes $b_1,\dots,
b_{M-1},a_0,\dots,a_{N-1}$, and the trace-preserving condition
requires that
\begin{eqnarray}
\left(\prod _{i=0}^{M-1} \int d^2 \beta _i \right ) 
\left(\bigotimes _{i=1}^{M-1} D_{b_i}
(\beta _i ) \right)U_{a_0b_0}(I_{b_0} \otimes C)U^\dag _{a_0b_0}
\left(\bigotimes _{i=1}^{M-1} D^\dag _{b_i}(\beta _i)\right)= I
\;,\label{tracep2}
\end{eqnarray}
which finally gives
\begin{equation}
\Tr_{b/b_0,a_0}[C]=\theta^2I_{a/a_0}\,.
\end{equation}
We now consider the map corresponding to
\begin{equation}
C=\theta^2 \,
|0\>\<0|_{b/b_0}\otimes|0\>\<0|_{a_0}\otimes 
I_{a/a_0}.
\label{ansconj}
\end{equation}
Applying such a map to a generic $N$-partite state $\Xi$ we get
\begin{equation}
\begin{split}
  \map C(\Xi)&=U_b\Tr_a[(I_b\otimes U^\dag_a\Xi^\tau  U_a)U_{a_0b_0}(I_{b_0}\otimes C)U^\dag_{a_0b_0}]U_b^\dag\\
  &=U_b(\Tr_{a_0}[(I_{b_0}\otimes\xi^\tau )U_{a_0b_0} (I_{b_0}\otimes
  |0\>\<0|_{a_0})U^\dag_{a_0b_0}]|0\>\<0|_{b/b_0})U_b^\dag,
\label{66}
\end{split}
\end{equation}
where $\xi^\tau =\Tr_{a/a_0}[U^\dag_a\Xi^\tau U_a]$. Moreover, one has 
\begin{equation}
  U_{a_0b_0}(I_{b_0}\otimes |0\>\<0|_{a_0})U^\dag_{a_0b_0}=
  \int\frac{\theta^2d^2\gamma}{\pi}|\eta\gamma\>\<\eta\gamma|_{b_0}\otimes|\theta\gamma\>\<\theta\gamma|_{a_0}\;,
\end{equation}
and Eq. (\ref{66}) gives 
\begin{equation}
  \map C(\Xi)=U_b\left(\map H(\xi)\otimes|0\>\<0|_{b/b_0}\right)U_b^\dag\,,
\end{equation}
where $\xi=\Tr_{a/a_0}[U_a\Xi U_a^\dag]$, and
\begin{equation}
  \map H(\rho)=\int\frac{d^2\gamma}\pi |(\eta/\theta)\gamma\>\<\gamma^*
  |\xi |\gamma^*\>\<(\eta/\theta)\gamma|.
\end{equation}
For $\Xi=\rho_0^{\otimes N}$ we have simply $\xi=\rho_0$, and this
implies that a simple scheme to achieve this map is the following.
First, the $N$ input states interact through an $N$-splitter, then the
system labeled $0$ carrying all the information about the coherent
signal is measured by heterodyne detection, and for any outcome 
$\gamma$ a coherent
state with amplitude $\sqrt{\frac MN}\gamma^*$ is generated. 
Finally, the prepared
state is sent through an $M$-splitter along with $M-1$ modes in the
vacuum state.\par

The output state $\map C(\rho_0^{\otimes N})$ is now given by
\begin{equation}
  \map C(\rho_0^{\otimes N})=\frac1{\bar n+1}\int \frac{d^2\gamma}\pi e^{-\frac{|\gamma|^2}{\bar n+1}}U_b(|\sqrt{M/N}\gamma\>\<\sqrt{M/N}\gamma|\otimes|0\>\<0|_{b/b_0})U_b^\dag,
\end{equation}
which is equal to
\begin{equation}
  \map C(\rho_0^{\otimes N})=\frac N{\bar n+1}\int \frac{d^2\gamma}\pi e^{-\frac{N|\gamma|^2}{\bar n+1}}|\gamma\>\<\gamma|,
\end{equation}
and its single-site reduced state is simply a thermal state with
\begin{equation}
  \bar n''=\frac{\bar n+1}N\,.
\end{equation}
Notice that this is independent of the number of output copies, and is
the same average number as the one for superbroadcasting in the limit
$M\to\infty$. More generally, the local output for generic input state
$\Xi$ is
\begin{equation}
  \rho ' = \int \frac{d^2 \alpha }{\pi}\,e ^{-\frac{|\alpha |^2}{2}(1+\frac
    1N)}\,
  \{\Tr [\Xi  D^\dag (\alpha  /N)^{\otimes N}] \} \, D(\alpha ).
\end{equation}\par

The proof of optimality is analogous to the proof for the
superbroadcasting map. It is sufficient to replace $U_{a_0b_0}$ with
$S_{a_0b_0}$ in Eqs.~\eqref{vutot} and \eqref{vuzel}.

\section{A proof of the optimality in terms of  linear amplifiers}
We are interested in a transformation that provides $M$ (generally
correlated) modes $b_0,b_1,...,b_{M-1}$ from $N$ uncorrelated modes
$a_0,a_1,...,a_{N-1}$, such that the unknown complex amplitude is
preserved and the output has minimal phase-insensitive noise. In
formula, we have input uncorrelated modes
\begin{eqnarray}
&& \< a_ i \>=\alpha \;, \nonumber \\& & 
\Delta x_{a_i} ^2 +\Delta y_{a_i} ^2 = \gamma_i \geq \frac 12, 
\label{}
\end{eqnarray}
for all $i=0,1,..,N-1$, where Heisenberg uncertainty relation is taken
into account. The output modes should satisfy
\begin{eqnarray}
&& \< b_ i \>=\alpha \;, \nonumber \\& & 
\Delta x_{b_i} ^2 +\Delta x_{b_i} ^2 = \Gamma \geq \frac 12
\;, \label{}
\end{eqnarray}
and we look for the minimal $\Gamma $. The minimal $\Gamma $ can be
obtained by applying a fundamental theorem for phase-insensitive
linear amplifiers \cite{caves}: the sum of the uncertainties of
conjugated quadratures of a phase-insensitive amplified mode with
(power) gain $G$ is bounded as follows.
\begin{eqnarray}
  \Delta X_B ^2 +\Delta Y_B ^2  \geq G
  (\Delta X_A ^2 +\Delta Y_A ^2  ) + \frac {G-1}{2}
  \;,\label{bou}
\end{eqnarray}
where $A$ and $B$ denotes the input and the amplified mode,
respectively. Our transformation can be seen as a phase-insensitive
amplification from the mode $A=\frac {1}{\sqrt N}\sum _{i=0}^{N-1}
a_i$ to the mode $B=\frac {1}{\sqrt M}\sum _{i=0}^{M-1} b_i$ with gain
$G=\frac MN$, and hence Eq. (\ref{bou}) should hold. Notice that
generally for any mode $c$ one has
\begin{eqnarray}
  \Delta x_c ^2 +\Delta y_c ^2  = \frac 12 + \< c^\dag c \> -|\<c\>|^2
  \;.\label{bou}
\end{eqnarray}
Hence, the bound can be rewritten as
\begin{eqnarray}
\< B^\dag B \> -|\<B\>|^2 \geq G(\< A^\dag A \> +1 -|\<A\>|^2) -1.
\end{eqnarray}
In the present case, since modes $a_i$ are uncorrelated, one has 
\begin{eqnarray}
&&\<A^\dag A\>= \frac 1N \sum _{i,j=0}^{N-1} \<a^\dag _i a_j\>= 
\frac 1N \left (\sum _{i=0}^{N-1} \<a^\dag _i a_i\>+ 
\sum _{i\neq j} \<a^\dag _i a_j\>\right )\nonumber \\& & 
= (\gamma + |\alpha |^2 -\frac 12)+
(N-1) |\alpha |^2 =\gamma +N |\alpha |^2 -\frac 12,
\end{eqnarray}
where $\gamma=\frac1N\sum_{i=0}^{N-1}\gamma_i$, and so the bound
Eq.~\eqref{bou} is written as
\begin{eqnarray}
\< B^\dag B \> 
\geq G (\gamma +\frac 12)-1+M |\alpha |^2.\label{ass}
\end{eqnarray}
On the other hand, one has 
\begin{eqnarray}
\<B^\dag B\>= \frac 1M \sum _{i,j=0}^{M-1} \<b^\dag _i b_j\>\leq 
\frac 1M \sum _{i,j=0}^{M-1} \sqrt{\<b^\dag _i b_i\> \<b^\dag _j b_j\> 
}= M(\Gamma + |\alpha |^2 -\frac 12).\label{ass2}
\end{eqnarray}
Eqs. (\ref{ass}) and (\ref{ass2}) together give the bound for the
minimal noise $\Gamma$ 
\begin{eqnarray}
\Gamma -\frac 12 \geq  \frac 1N (\gamma -\frac 12) +\frac {1}{N}-
\frac {1}{M}.
\end{eqnarray}
The example in the previous sections corresponds to $\gamma = {\bar
  n}+ \frac 12$ and $\Gamma ={\bar n ''}+\frac 12$. A similar
derivation gives a bound for purification, where $N > M$.  In such a
case $G<1$, and Eq. (\ref{bou}) is replaced with
\begin{eqnarray}
\Delta X_B ^2 +\Delta Y_B ^2  \geq G
(\Delta X_A ^2 +\Delta Y_A ^2  ) + \frac {1-G}{2}
\;,\label{bou2}
\end{eqnarray}
and one obtains the bound
\begin{eqnarray}
\Gamma -\frac 12 \geq  \frac 1N (\gamma -\frac 12) 
\;.\label{pur}
\end{eqnarray}
We would like to stress that the derivation of all bounds in the
present section relies on the theorem of the added noise in {\em
  linear} amplifiers, namely only linear transformations of modes are
considered.  Hence, in principle, these bounds might be violated by
more exotic and nonlinear transformations. Therefore, the derivation
of Eq.  (\ref{bnd}) is stronger, since it has general validity.\par

By a similar derivation, using the bound for phase-conjugated
amplifiers $\Delta X_B^2+\Delta Y_B^2\geq G(\Delta X_A^2+\Delta
Y_A^2)+\frac{G-1}2$, one can obtain the bound for phase-conjugation
broadcasting 
\begin{equation}
\Gamma-\frac12\geq\frac1N\left(\gamma+\frac12\right).
\end{equation}

\section{Experimental implementation}
\begin{figure}[h]
\epsfig{file=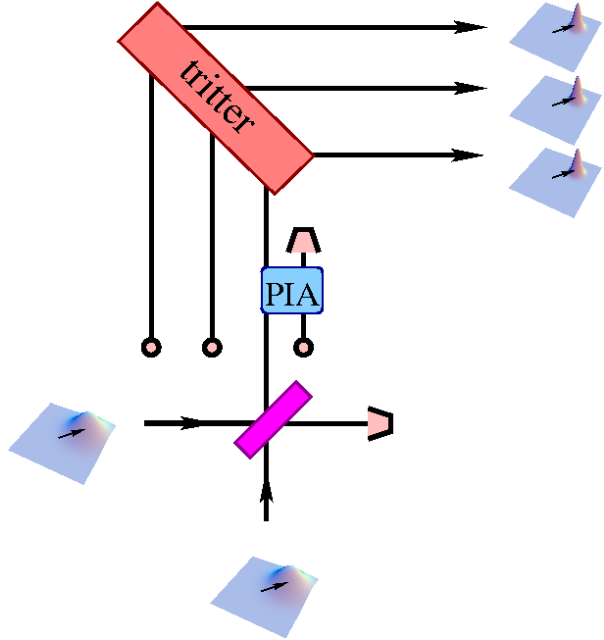,width=8cm}
\caption{Experimental scheme to achieve optimal superbroadcasting from
  2 to 3 copies. This setup involves just a beam splitter, a
  phase-insensitive amplifier and a tritter, which in turn can be
  implemented by two suitably balanced beam splitters. The
  phase-insensitive amplifier can be implemented by a beam splitter
  and heterodyne-assisted feed-forward. The output copies carrying the
  same signal as the input ones are locally more pure, the noise being
  shifted to classical correlations between them.\label{sch}}
\end{figure}

The optimal broadcasting can be easily implemented by means of an
inverse $N$-splitter which concentrates the signal in one mode and
discards the other $N-1$ modes. The mode is then amplified by a
phase-insensitive amplifier with power gain $G=\frac MN$. Finally, the
amplified mode is distributed by mixing it in an $M$-splitter with
$M-1$ vacuum modes. Each mode is then found in the state of Eq.
(\ref{rhogen}). In the concentration stage the $N$ modes with
amplitude $\<a_i\>=\alpha $ and noise $\Delta x_i^2 +\Delta y_i^2
=\gamma_i$ are reduced to a single mode with amplitude $\sqrt N \alpha
$ and noise $\gamma $. The amplification stage gives a mode with
amplitude $\sqrt M \alpha $ and noise $\gamma ' =\gamma \frac MN +
\frac {M}{2N} -\frac 12$.  Finally, the distribution stage gives $M$
modes, with amplitude $\alpha $ and noise $\Gamma = \frac 1M \left
  (\gamma ' +\frac {M-1}{2}\right )$ each. In Fig. \ref{sch} we sketch
the scheme for $2$ to $3$ superbroadcasting.\par

In Ref. \cite{leuchs} it was shown experimentally that phase
insensitive amplification can be obtained by a setup consisting of a
beam-splitter, a heterodyne detector and a conditional displacement.
In the following we give an algebraic derivation of this result.
Consider a mode in a state $\rho=\int\frac{d^2\gamma}\pi
f(\gamma)D(\gamma)$ coupled to another mode in the vacuum through a
beam-splitter with transmissivity $\tau $. The output is given by the
bipartite state $\sigma$
\begin{equation}
\sigma=\int\frac{d^2\beta d^2\gamma}{\pi^2}e^{-\frac{|\tau \beta-\sqrt{1-\tau ^2}\gamma|^2}2} f(\tau \gamma+\sqrt{1-\tau ^2}\beta)D(\gamma)\otimes D(\beta),
\end{equation}
where we performed the change of variables
$\beta\to\tau\beta+\sqrt{1-\tau^2}\gamma$,
$\gamma\to\tau\gamma-\sqrt{1-\tau^2}\beta$. Now, the reflected mode is
measured by heterodyne detection, and conditionally on the measurement
outcome $\alpha$, a displacement $D(k\alpha)$ is performed on the
transmitted mode, whose state is then given by
\begin{align}
  \rho'= &\int\frac{d^2\alpha d^2\beta d^2\gamma}{\pi^3}e^{-\frac{|\tau\beta-\sqrt{1-\tau^2}\gamma|^2}2} f(\tau\gamma+\sqrt{1-\tau^2}\beta)D(k\alpha)D(\gamma)D(k\alpha)^\dag\<0|D(\alpha)^\dag D(\beta)D(\alpha)|0\>=\nonumber\\
  &\int\frac{d^2\alpha d^2\beta d^2\gamma}{\pi^3}e^{-\frac{|\tau\beta-\sqrt{1-\tau^2}\gamma|^2}2} f(\tau\gamma+\sqrt{1-\tau^2}\beta)e^{\alpha(k\gamma^*-\beta^*)-c.c.}D(\gamma)e^{-\frac{|\beta|^2}2}=\nonumber\\
  &\int\frac{d^2\beta d^2\gamma}{\pi^2}\delta^{(2)}(k\gamma-\beta)e^{-\frac{|\tau\beta-\sqrt{1-\tau^2}\gamma|^2}2} f(\tau\gamma+\sqrt{1-\tau^2}\beta)e^{-\frac{|\beta|^2}2}=\nonumber\\
  &\int\frac{d^2\gamma}{\pi}f(\gamma(\tau+k\sqrt{1-\tau^2}))e^{-\frac{|\gamma|^2}2[k^2+(k\tau-\sqrt{1-\tau^2})^2]}D(\gamma).
\end{align}
On the other hand, the action of a phase-insensitive amplifier on
$\rho$ can be easily calculated and produces the partial output state
\begin{equation}
\rho''=\int\frac{d^2\gamma}{\pi}e^{-\frac{|\gamma|^2\nu^2}2}f(\mu\gamma)D(\gamma).
\end{equation}
The following conditions
\begin{equation}
  \mu=\tau+k\sqrt{1-\tau^2},\quad\nu^2=k^2+(k\tau-\sqrt{1-\tau^2})^2,\quad\mu^2-\nu^2=1,
\end{equation}
which equivalent to
\begin{equation}
  k=\nu,\quad\tau=\frac1\mu,
\end{equation}
imply that $\rho'=\rho''$.Hence, by tuning the beam splitter
transmissivity and the parameter of the conditional displacement $k$,
one can then simulate the amplifier by a linear device assisted by
heterodyne and feed-forward.\par

The optimal phase-conjugated broadcasting can be obtained by replacing
the linear amplifier with a heterodyne measurement and preparation
of a coherent state with conjugate phase and amplified intensity. For
achieving the optimal purification, one simply uses an inverse
$N$-splitter which concentrates the signal in one mode and discards
the other $N-1$ modes. Then by $N$-splitting with $N-1$ vacuum modes,
one obtained $N$ purified signals (although classically correlated).

\section{Conclusion}
\label{concl}

In conclusion, we proved that broadcasting of $M$ copies of a mixed
radiation state starting from $N<M$ copies is possible, even with
lowering the total noise in conjugate quadratures. Since the noise
cannot be removed without violating the quantum data processing
theorem, the price to pay for having higher purity at the output is
that the output copies are correlated. Essentially noise is moved from
local states to their correlations, and our superbroadcasting channel
does this optimally.  We obtained similar results also for
purification (i.e.  $M\leq N$), along with the case of simultaneous
broadcasting and phase-conjugation, with the output copies carrying a
signal which is complex-conjugated of the input one.  Despite the role
that correlations play in this effect, no entanglement is present in
the output (as long as the single input copy has a positive
$P$-function), as it can be seen by the analytical expression of the
output states.  Moreover, a practical and very simple scheme for
experimental achievement of the maps has been shown, involving mainly
passive media and only one parametric amplifier. The superbroadcasting
effect has a relevance form the fundamental point of view, opening new
perspectives in the understanding of correlations and their interplay
with noise, but may be also promising from a practical point of view,
for communication tasks in the presence of noise.

\end{document}